\documentclass[aps,twocolumn,superscriptaddress]{revtex4}
\usepackage{amsfonts}
\usepackage{amsmath}
\usepackage{amssymb}
\usepackage{graphicx}
\usepackage{xcolor}
\usepackage{float}
\usepackage{lipsum}

\begin{document}
\title{Spatial Kramers-Kronig relation and unidirectional light reflection induced by Rydberg dipole-dipole interactions}
\author{Di-Di Zheng}
\affiliation{School of Physics and Center for Quantum Sciences,
Northeast Normal University, Changchun 130024, China}
\author{Yan Zhang}
\affiliation{School of Physics and Center for Quantum Sciences,
Northeast Normal University, Changchun 130024, China}
\author{Yi-Mou Liu}
\affiliation{School of Physics and Center for Quantum Sciences,
Northeast Normal University, Changchun 130024, China}
\author{Xiao-Jun Zhang}
\email{zhangxj037@nenu.edu.cn} \affiliation{School of Physics and
Center for Quantum Sciences, Northeast Normal University, Changchun
130024, China}
\author{Jin-Hui Wu}
\email{jhwu@nenu.edu.cn} \affiliation{School of Physics and Center
for Quantum Sciences, Northeast Normal University, Changchun 130024,
China}
\date{\today }

\begin{abstract}
Kramers-Kronig (KK) relation between the dispersion and absorption
responses of a signal field can be mapped from the frequency domain
into the space domain via the dipole-dipole interactions between a
homogeneous sample of target atoms and a control atom. This is
achieved by establishing an effective two-level configuration for
the three-level target atoms in the single-photon far-detuned
driving regime while maintaining a high Rydberg excitation for the
three-level control atom in the single-photon resonant driving
regime. We find in particular that it is viable to realize a
dynamically tunable spatial KK relation supporting asymmetric and
even unidirectional reflection for appropriate signal frequencies in
a controlled range. Taking a periodic lattice of target atoms
instead, multiple Bragg scattering can be further incorporated into
spatial KK relation to largely enhance the nonzero reflectivity yet
without breaking the asymmetric or unidirectional reflection.
\end{abstract}
\maketitle

\section{Introduction}
In recent years, great efforts have been made in the realization and
manipulation of asymmetric light reflection and even unidirectional
invisibility with artificial optical structures of complex optical
potentials~\cite{Lin.2011,Longhi.2011,Regensburger.2012,Mostafazadeh.2013,Feng.2013,Castaldi.2013,Fu.2016,Rivolta.2016,Liu.2017,Huang.2017,Sarsaman.2017,Sarsaman.2018,Yuan.2019,Horsley.2015,Longhi.2016,King.2017,Horsley.2017,Horsley.2017a,Longhi.2017,DLiu.2020,Jiang.2017,Ye.2017,Singh.2020,Zhang.2021}.
One main motivation lies in that relevant advances are essential for
developing one-way optical devices unattainable with natural linear
materials of real optical potentials. Reflection and transmission
properties are typically bidirectional and symmetric for isotropic
linear materials based on the Lorentz reciprocal
theorem~\cite{Haus.1984,Saha.2017}. This can also be understood in
view of information optics, which argues that the Fourier transform
of a real optical potential is definitely symmetric so that light
propagation in natural linear materials always results in balanced
forward and backward modes~\cite{Kulishov.2005,Yang.2016}.

Now it is known that unidirectional reflection and invisibility can
be attained at an exceptional point in non-Hermitian media
exhibiting, \textit{e.g.}, parity-time (PT)
symmetry~\cite{Lin.2011,Longhi.2011,Regensburger.2012,Mostafazadeh.2013,Feng.2013,Castaldi.2013,Fu.2016,Rivolta.2016,Liu.2017,Huang.2017,Sarsaman.2017,Sarsaman.2018,Yuan.2019}.
These media are, however, very challenging in regard of the
experimental implementation because they require elaborate designs
of gain and loss. Horsley \textit{et al.} found in 2015 that
electromagnetic waves incident upon an inhomogeneous medium, the
real and imaginary parts of whose complex permittivity are related
in space via the Kramers-Kronig (KK) relation, can be efficiently
absorbed from one side but normally reflected from the other
side~\cite{Horsley.2015}. Soon afterwards, results in this pioneer
work were extended in
theory~\cite{Longhi.2016,King.2017,Horsley.2017,Horsley.2017a,Longhi.2017,DLiu.2020},
verified in experiment~\cite{Jiang.2017,Ye.2017,Singh.2020}, and
explored to develop new techniques of holographic imaging or
anechoic chamber~\cite{Baek.2021,Lee.2022,QLi.2022}. Such spatial KK
media, though requiring no elaborate designs on gain and loss, are
typically designed with fixed structures and lack the dynamic
tunability. A feasible method for overcoming this difficulty is to
consider multi-level driven atomic systems, in which the
frequency-to-space mapping of an induced susceptibility can be
attained via a dynamic Stark or Zeeman effect~\cite{Zhang.2021}.

On the other hand, we note that nonlocal dipole-dipole interactions
(DDIs) of Rydberg atoms depend critically on the interatomic
distance $R$ and can be manipulated on demand by external driving
fields~\cite{Tong.2004,Vogt.2006,Baluktsian.2013,Fan.2020}. This
then motivates us to seek a feasible driving scheme where DDIs can
be used to realize the spatial KK relation by establishing a
nonlinear dependence of atomic transition frequency on atomic
spatial position. To be more specific, DDIs may manifest as either
van der Waals ($vdW$) potentials scaling as $1/R^{6}$ in the
non-resonant regime or F\"{o}rster-like potentials scaling as
$1/R^{3}$ in the resonant regime~\cite{Nguyen.2016}. In fact,
Rydberg atoms have been well studied as an intriguing platform for
realizing quantum information
processing~\cite{Lukin.2001,Barato.2014,Bernien.2017,Levine.2018,Adams.2021}
and high-precision field
sensing~\cite{Sedlacek.2013,Fan.2015,Wade.2018,Cox.2018,Wang.2020},
considering that they also exhibit the features of long radiative
lifetimes and large electric dipole moments. Note also that Rydberg
atoms have been explored in the regime of electromagnetically
induced transparency (EIT) to achieve effective interactions between
individual photons, which promise the realization of nontrivial
photonic devices like single-photon
sources~\cite{saffman.2002,Peyronel.2012,Rapika.2018},
memorizers~\cite{Li.2016,Distante.2017,Zhangh.2021}, and
transistors~\cite{Gorniaczyk.2014,Tiarks.2014,Hao.2019}. To the best
of our knowledge, DDIs of Rydberg atoms have not been considered to
develop photonic devices supporting asymmetric light propagation
behaviors.

\vspace{2mm} We examine here an effective scheme for the realization
of a tunable spatial KK relation in a homogeneous sample of cold
target atoms by utilizing their $vdW$ interactions with a control
atom. This is done by considering a single-photon resonant driving
configuration for the control atom in the
dark-state~\cite{Gray:78,quantum-optics} regime while a
single-photon far-detuned driving configuration for the target atoms
in the EIT regime. Under appropriate conditions, the control atom
can be made to exhibit a roughly perfect Rydberg excitation via a
dark-state manipulation while the target atoms may reduce from a
three-level to a two-level configuration by adiabatically
eliminating the intermediate state. On this account, it is viable to
realize a nonlinear frequency-to-space mapping of the dispersion and
absorption responses and hence a well established and modulated
spatial KK relation. Consequently, the reflectivity of a signal
field incident upon one side is distinct from that upon the other
side and may even become vanishing to result in unidirectional
reflection. Replacing the homogeneous atomic sample with a periodic
atomic lattice, we further show it is viable to improve the
asymmetric and unidirectional reflection behaviors, by largely
enhancing the nonzero reflectivity yet without activating the
vanishing reflectivity, when multiple Bragg scattering is
incorporated into spatial KK relation.

\section{Model and Equations}

\begin{figure}[tb]
\centering \includegraphics[width=0.48\textwidth]{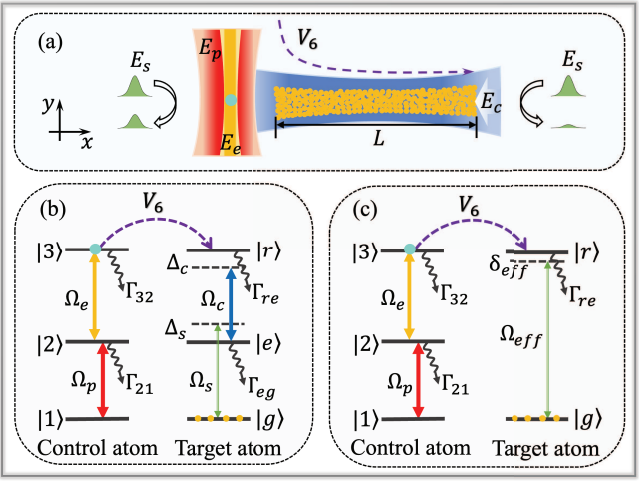}
\caption{(Color online) (a) Schematic of asymmetric reflection of a
signal ($E_{s}$) field incident upon a homogeneous sample of target
atoms extending from $x=0$ to $x=L$ in the presence of a coupling
($E_{c}$) field. A control atom irradiated by a pumping ($E_{p}$)
and an exciting ($E_{e}$) field is placed at $x=x_{0}$ to manipulate
all target atoms via $vdW$ ($\mathcal{V}_{6}$) interactions. (b)
Driving configurations for a pair of control and target atoms
interacting via a $\mathcal{V}_{6}$ potential. The pumping
($\Omega_{p}$) and exciting ($\Omega_{p}$) fields are on both
single-photon and two-photon resonances with relevant transitions of
the control atom. The signal ($\Omega_{s}$) and coupling
($\Omega_{c}$) fields are far-off single-photon resonance but
meanwhile near two-photon resonance with relevant transitions of the
target atom. (c) Effective configurations for a pair of control and
target atoms when the intermediate state $\vert e \rangle$ is
eliminated and the signal and coupling fields are replaced by an
effective ($\Omega_{eff}$) field under appropriate conditions.}
\label{fig:Fig1}
\end{figure}

We start by introducing our basic model in Fig.~\ref{fig:Fig1}(a),
where a signal field of amplitude (frequency) $E_{s}$ ($\omega_{s}$)
is incident upon a homogeneous sample of cold target atoms from the
$x=0$ or $x=L$ side, and the overall optical response of target
atoms is modulated by a control atom at $x=x_{0}$ via $vdW$
interactions relevant to a high Rydberg excitation. The control atom
is driven by a pumping field of amplitude (frequency) $E_{p}$
($\omega_{p}$) on transition $|1\rangle\leftrightarrow|2\rangle$ and
an exciting field of amplitude (frequency) $E_{e}$ ($\omega_{e}$) on
transition $|2\rangle\leftrightarrow|3\rangle$ as shown in
Fig.~\ref{fig:Fig1}(b), being $\Omega_{p}=E_{p}\wp_{21}/2\hbar$ and
$\Omega_{e}=E_{e}\wp_{32}/2\hbar$ corresponding Rabi frequencies
while $\Delta_{p}=\omega_{p}-\omega_{21}$ and
$\Delta_{e}=\omega_{e}-\omega_{32}$ corresponding detunings. The
target atoms are driven instead by the signal field on transition
$|g\rangle\leftrightarrow|e\rangle$ and a coupling field of
amplitude (frequency) $E_{c}$ ($\omega_{c}$) on transition
$|e\rangle\leftrightarrow|r\rangle$ as shown in
Fig.~\ref{fig:Fig1}(b), being $\Omega_{s}=E_{s}\wp_{eg}/2\hbar$ and
$\Omega_{c}=E_{c}\wp_{re}/2\hbar$ corresponding Rabi frequencies
while $\Delta_{s}=\omega_{s}-\omega_{eg}$ and
$\Delta_{c}=\omega_{c}-\omega_{re}$ corresponding detunings. Above
we have used $\wp_{\mu\nu}$ and $\omega_{\mu\nu}$ to denote dipole
moments and resonant frequencies, respectively, on transitions $|\mu
\rangle \leftrightarrow |\nu \rangle$ with $\{\nu,\mu\}\in\{1,2,3\}$
for the control atom while $\{\nu,\mu\}\in\{g,e,r\}$ for the target
atoms. In addition $\Delta_{p}=\Delta_{e}=0$ and
$\Delta_{s}\simeq-\Delta_{c}$ have been considered in
Fig.~\ref{fig:Fig1}(b) as an illustration of our interest.

It is worth noting that, the signal and coupling fields have
negligible effects on, despite traveling through, the control atom
because they are assumed to be far detuned from the $\vert 1 \rangle
\leftrightarrow \vert 2 \rangle$ and $\vert 2 \rangle
\leftrightarrow \vert 3 \rangle$ transitions, respectively. This may
be achieved by considering ground states $\vert 1 \rangle\equiv
\vert 5S_{1/2},F=1\rangle$ and $\vert g \rangle \equiv \vert
5S_{1/2},F=2\rangle$, intermediate states $\vert 2\rangle\equiv
\vert 5P_{3/2},F=0\rangle$ and $\vert e \rangle \equiv \vert
5P_{3/2},F=3\rangle$, and Rydberg states $\vert 3 \rangle=\vert r
\rangle \equiv \vert 90S_{1/2}\rangle$ for the $^{87}$Rb isotope.
The pumping and exciting fields, however, don't travel through the
target atoms as arranged in Fig.~\ref{fig:Fig1}(a). With above
considerations, we can easily write down the following Hamiltonians
by setting $\hbar=1$
\begin{align}
H_{c}&= -\Delta_{p}\sigma_{22}-(\Delta_{p}
+\Delta_{e})\sigma_{33}-\Omega_{p}\sigma_{21}-\Omega_{e}\sigma_{32}\notag \nonumber\\
&\qquad-\Omega_{p}^{\ast}\sigma_{12}-\Omega_{e}^{\ast}\sigma_{23},\label{Eqn1}\\
H_{t}&=
-\Delta_{s}\sigma_{ee}-(\Delta_{s}+\Delta_{c})\sigma_{rr}-\Omega_{s}\sigma_{eg}-\Omega_{c}\sigma_{re}\nonumber\\
&\qquad-\Omega_{s}^{\ast}\sigma_{ge}-\Omega_{c}^{\ast}\sigma_{er}+\mathcal{V}_{6}\sigma_{33}\sigma_{rr},\nonumber
\end{align}
for the control and target atoms in order. Here we introduce
$\sigma_{\nu\mu}=\vert\nu\rangle\langle\mu\vert$ to denote the
transition ($\nu\ne\mu$) or projection ($\nu=\mu$) operator, while
$\mathcal{V}_{6}=C_{6}/(x-x_{0})^{6}$ represents the $vdW$ potential
of coefficient $C_{6}$ for the control atom at $x_{0}<0$ and a
target atom at $x\geq 0$.

Dynamic evolution of the control atom is governed by the master
equation for density operator $\rho$
\begin{align}
\partial_{t}\rho=-i[H_{c},\rho]+\mathcal{L}_{c}(\rho),\label{Eqn2}
\end{align}
where
$\mathcal{L}_{c}(\rho)=\sum\Gamma_{\mu\nu}[\sigma_{\nu\mu}\rho\sigma_{\mu\nu}-\frac
{1}{2}(\rho\sigma_{\mu\nu}\sigma_{\nu\mu}+\sigma_{\mu\nu}\sigma_{\nu\mu}
\rho)]$ describes the dissipation processes contributed by
population decay rates $\Gamma_{32}$ and $\Gamma_{21}$ on the $\vert
3\rangle\leftrightarrow\vert 2\rangle$ and $\vert
2\rangle\leftrightarrow\vert 1\rangle$ transitions, respectively.
Using $H_{c}$ and $\mathcal{L}_{c}(\rho)$, it is easy to expand
Eq.~(\ref{Eqn2}) into a set of dynamic equations on nine density
matrix elements $\rho_{\mu\nu}$ with $\{\mu,\nu\}\in\{1,2,3\}$.
These equations can be solved by setting
$\partial_{t}\rho_{\mu\nu}=0$ to attain the
dark-state~\cite{Gray:78,quantum-optics} Rydberg population
\begin{align}
\rho_{33}\simeq\frac{(\gamma_{21}+\gamma_{31})\Omega_{p}^{2}\Omega_{e}^{2}}{\gamma_{21}\Omega_{e}^{4}
+(\gamma_{21}+3\gamma_{31})\Omega_{p}^{2}\Omega_{e}^{2}+\gamma_{21}^{2}\gamma_{31}\Omega_{e}^{2}},\label{Eqn3}
\end{align}
in the limit of $\Delta_{p}=\Delta_{e}=0$ and
$\Omega_{p}\geq\Omega_{e}>\gamma_{21}\gg\gamma_{31}$ with
$\gamma_{31}=\Gamma_{32}/2+\gamma_{31}^{d}$ and
$\gamma_{21}=\Gamma_{21}/2$. Here $\gamma_{31}^{d}$ denotes a pure
dephasing rate arising from finite laser linewidths and has to be
included because $\Gamma_{32}$ is negligible for high Rydberg
states. Moreover, keep in mind that $\gamma_{31}$ should be much
smaller than other parameters so as to maintain the dark state
$\vert D\rangle=c_{1}\vert 1 \rangle-c_{3}\vert 3 \rangle$ by
excluding state $\vert 2 \rangle$, hence it is viable to attain
$\rho_{33}=|c_{3}|^{2}\simeq\Omega_{p}^2/(\Omega_{p}^{2}+\Omega_{e}^{2})\to
1$ of our interest by further requiring
$\Omega_{p}^{2}\gg\Omega_{e}^{2}$.

With the same strategy, after introducing population decay rates
$\Gamma_{re}$ and $\Gamma_{eg}$ as well as dephasing rates
$\gamma_{re}=(\Gamma_{re}+\Gamma_{eg})/2$,
$\gamma_{rg}=\Gamma_{re}/2+\gamma_{rg}^{d}$, and
$\gamma_{eg}=\Gamma_{eg}/2$, we can write down a new set of dynamic
equations on nine density matrix elements $\rho_{\mu\nu}$ with
$\{\mu,\nu\}\in\{g,e,r\}$ for the target atoms. These equations can
be solved by setting $\partial_{t}\rho_{\mu\nu}=0$ and $\rho_{ee}\to
0$ in the limit of $\Delta_{s}\simeq-\Delta_{c}$,
$|\Delta_{s}|\gg\gamma_{eg}\gg\Omega_{s}$, and
$|\Delta_{c}|\gg\Omega_{c}\gg\gamma_{re}$ to attain
\begin{align}
\rho_{gg}&=\frac{\Gamma_{re}[\gamma_{rg}^{2}+(\delta_{eff}+\mathcal{V}_{6}\rho_{33})^{2}]+2\gamma_{rg}\Omega_{eff}^{2}}
{\Gamma_{re}[\gamma_{rg}^{2}+(\delta_{eff}+\mathcal{V}_{6}\rho_{33})^{2}]+4\gamma_{rg}\Omega_{eff}^{2}},\label{Eqn4}\\
\rho_{rg}&=\frac{i\Omega_{eff}\Gamma_{re}[\gamma_{rg}+i(\delta_{eff}+\mathcal{V}_{6}\rho_{33})]}
{\Gamma_{re}[\gamma_{rg}^{2}+(\delta_{eff}+\mathcal{V}_{6}\rho_{33})^{2}]+4\gamma_{rg}\Omega_{eff}^{2}},\nonumber
\end{align}
restricted by
$\rho_{eg}=-(\Omega_{c}^{\ast}\rho_{rg}+\Omega_{s}\rho_{gg})/\Delta_{s}$,
$\rho_{re}=(\Omega_{s}^{\ast}\rho_{rg}+\Omega_{c}\rho_{rr})/\Delta_{c}$,
and $\rho_{gg}+\rho_{rr}=1$. Here
$\Omega_{eff}=\Omega_{s}\Omega_{c}/\Delta_{c}$ is an effective
two-photon Rabi frequency while
$\delta_{eff}=\Delta_{s}+\Delta_{c}-\Delta_{e1}-\Delta_{e2}$ is an
effective two-photon detuning modified by
$\Delta_{e1}=\Omega_{c}^{2}/\Delta_{s}$ and
$\Delta_{e2}=\Omega_{s}^{2}/\Delta_{c}$.

Further considering $\gamma_{rg}\Gamma_{re}\gg4\Omega_{eff}^{2}$,
which is available by enhancing $\gamma_{rg}$ with finite laser
linewidths~\cite{YQLi.1995} and $\Gamma_{re}$ via incoherent
(downward) pumpings~\cite{SWDu.2017}, we can attain with
Eq.~(\ref{Eqn4}) an induced signal susceptibility
\begin{align}
\chi_{s}
=\frac{N_{0}\wp_{ge}^{2}}{\hbar\varepsilon_{0}}\left[\frac{\Omega_{c}^{2}}{\Delta_{s}\Delta_{c}}
\frac{\delta_{eff}+\mathcal{V}_{6}\rho_{33}-i\gamma_{rg}}{\gamma_{rg}^{2}+(\delta_{eff}+\mathcal{V}_{6}\rho_{33})^{2}}-\frac{1}{\Delta_{s}}\right],\label{Eqn5}
\end{align}
describing the target atoms reduced to a two-level configuration as
shown in Fig.~\ref{fig:Fig1}(c). It is worth noting that $\chi_{s}$
is position-dependent in the presence of a $vdW$ potential
$\mathcal{V}_{6}$ and valid only in the case of
$|\delta_{eff}|\ll|\Delta_{s}\simeq-\Delta_{c}|$. We also note that
the real ($\chi_{s}^{\prime}$) and imaginary
($\chi_{s}^{\prime\prime}$) parts of $\chi_{s}$ describe,
respectively, the dispersion and absorption responses and are
connected via the KK relation in the frequency domain based on the
causality principle and Cauchy's theorem in the case of
$\mathcal{V}_{6}=0$~\cite{Landau.1984}.

The KK relation may also hold in the space domain in the case of
$\mathcal{V}_{6}\ne 0$ for appropriate values of $\delta_{eff}$.
This is true only if $\chi_{s}^{\prime}$ and
$\chi_{s}^{\prime\prime}$ are related through
\begin{equation}
\chi_{s}^{\prime}(\delta_{eff},
x)=\frac{1}{\pi}\texttt{P}\int_{0}^{L}
\frac{\chi_{s}^{\prime\prime}(\delta_{eff},\xi)}{\xi-x}
d\xi,\label{Eqn6}
\end{equation}
where $\texttt{P}$ denotes a Cauchy's principle-value integral with
respect to atomic position $\xi$. Eq.~(\ref{Eqn6}) indicates that
$\chi_{s}^{\prime}$ and $\chi_{s}^{\prime\prime}$ must be spatially
out of phase in the case of a perfect spatial KK relation such that
the target atoms becomes unidirectional reflectionless to the signal
field~\cite{Horsley.2015}. This can be understood by considering
that, if $\chi_{s}^{\prime}$ and $\chi_{s}^{\prime\prime}$ are
spatially out of phase and meanwhile analytical in the upper half
complex plane, their Fourier components contain only positive
wavevectors and hence give rise to no backscattering relevant to
negative wavevectors. The degree to which the spatial KK relation is
violated can be evaluated by a figure of merit defined as
\begin{equation}
D_{kk}=\frac{\int_{0}^{L}{[\chi_{s}^{\prime\prime}(\delta_{eff},x)-\frac{1}{\pi}\texttt{P}\int_{0}^{L}
\frac{\chi_{s}^{\prime}(\delta_{eff},\xi)}{\xi-x}d\xi}]dx}{\int_{0}^{L}\chi_{s}^{\prime\prime}(\delta_{eff},x)dx}.\label{Eqn7}
\end{equation}
Consequently, $D_{kk}=0$ denotes a perfect spatial KK relation in
the unbroken regime while a larger $|D_{kk}|$ indicates a greater
degree of violation in the broken regime.

To examine the reflection and transmission spectra, we resort to the
transfer matrix method~\cite{Artoni.2006} sketched below. First, we
partition the atomic sample into a large number ($J\gg1$) of thin
slices labeled by indices $j\in\{1,J\}$, which exhibit an identical
thickness $\ell=L/J$ but different susceptibilities
$\chi_{s}(\delta_{eff},x)\to\chi_{s}(\delta_{eff},j\ell)$ . Second,
we establish a $2\times 2$ unimodular transfer matrix
$M_{j}(\delta_{eff},\ell)$ with $\chi_{s}(\delta_{eff},j\ell)$ to
describe the propagation of a signal field of wavelength
$\lambda_{s}$ through the $j$th slice via
\begin{equation}
\left[\begin{array}{c}
E^{+}_s(\delta_{eff},j\ell)\\
E^{-}_s(\delta_{eff},j\ell)
\end{array}
\right] =M_j(\delta_{eff},\ell) \left[\begin{array}{c}
E^{+}_s(\delta_{eff},j\ell-\ell)\\
E^{-}_s(\delta_{eff},j\ell-\ell)
\end{array}
\right],\label{Eqn8}
\end{equation}
where $E^{+}_{s}$ and $E^{-}_{s}$ denote, respectively, the forward
and backward components of a scattered signal field. Third, it is
straightforward to attain the total transfer matrix
$M(\delta_{eff},L)=M_{J}(\delta_{eff},\ell)\cdot\cdot\cdot
M_{j}(\delta_{eff},\ell)\cdot\cdot\cdot M_{1}(\delta_{eff},\ell)$ as
a sequential multiplication of the individual transfer matrices of
all slices of the atomic sample. Finally, we can write down the
(\textit{asymmetric}) reflectivities $R_{l}\ne R_{r}$ and
(\textit{reciprocal}) transmissivities $T=T_{l,r}$ in terms of
relevant matrix elements $M_{(ij)}(\delta_{eff},L)$ as given by
\begin{align}
R_l(\delta_{eff},L) & =|r_l(\delta_{eff},L)|^2=\left|\frac{M_{(12)}(\delta_{eff},L)}{M_{(22)}(\delta_{eff},L)}\right|^2,\nonumber\\
R_r(\delta_{eff},L) & =|r_r(\delta_{eff},L)|^2=\left|\frac{M_{(21)}(\delta_{eff},L)}{M_{(22)}(\delta_{eff},L)}\right|^2,\label{Eqn9}\\
T(\delta_{eff},L) &
=|t(\delta_{eff},L)|^2=\left|\frac{1}{M_{(22)}(\delta_{eff},L)}\right|^2,\nonumber
\end{align}
where `$l$' and `$r$' refer to a signal field incident from the left
($x=0$) and right ($x=L$) sides, respectively.

So far we have been considering a homogeneous sample of cold target
atoms. Now we switch to another scenario where target atoms are
trapped in an optical lattice of period $\Lambda$ and exhibit a
periodic Gaussian density
\begin{equation}
N(x)=\sum\limits_{k=1}^{K}N_{k}(x)=\sum\limits_{k=1}^{K}
\frac{\Lambda N_{0}}{\delta x \sqrt{\pi}} e^{-(x-x_{k})^{2}/\delta
x^{2}}.\label{Eqn10}
\end{equation}
Here $x_{k}=(k-1/2)\Lambda$ denotes the $k$th unit cell's center
while $\delta x$ and $\Lambda N_{0}/\delta x \sqrt{\pi}$ are the
common width and peak of all unit cells, respectively. This atomic
lattice of mean density $N_{0}$ and length $L=K\Lambda$ will be
examined to show how a nonzero reflectivity is enhanced by
incorporating multiple Bragg scattering into spatial KK relation.

\section{Results and Discussion}

We now begin to examine the out-of-phase spatial distributions of
$\chi_{s}^{\prime}$ and $\chi_{s}^{\prime\prime}$ as well as the
asymmetric spectra of $R_{l}$ and $R_{r}$ with formulas developed in
the last section. To this end, we first specify realistic parameters
for the states of $^{87}$Rb isotopes mentioned before
Eq.~(\ref{Eqn1}) with $\Gamma_{32}/2\pi=0.5$ kHz,
$\Gamma_{re}/2\pi=40$ kHz, $\Gamma_{21,eg}/2\pi=6.0$ MHz,
$\gamma_{31,rg}^{d}/2\pi=20$ kHz, $\wp_{eg}=2.54\times 10^{-29}$
C$\cdot$m, and $C_{6}/2\pi=1.68\times10^{13}$ $s^{-1}\mu
m^{6}$~\cite{Singer.2005,Beterov.2009,Steck.2021}. With respect to
the applied fields, we may further choose $\Omega_{p}/2\pi=50$ MHz,
$\Omega_{e}/2\pi=5.0$ MHz, and $\Delta_{p}=\Delta_{e}=0$ to achieve
a high enough Rydberg population ($\rho_{33}\to 1$) for the control
atom, while $\Omega_{s}/2\pi=40$ kHz, $\Omega_{c}/2\pi=10$ MHz, and
$-\Delta_{s}/2\pi\simeq \Delta_{c}/2\pi=200$ MHz to justify the
two-level approximation ($\Omega_{eff}/2\pi=2.0$ kHz,
$\Delta_{e1}/2\pi=-0.5$ MHz, and $\Delta_{e2}\to 0$) for all target
atoms.

\begin{figure}[tb]
\centering \includegraphics[width=0.48\textwidth]{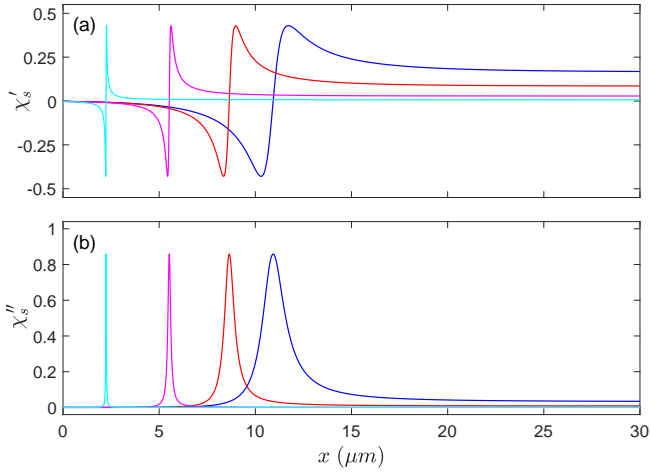}
\caption{(Color online) (a) Real and (b) imaginary parts of signal
susceptibility $\chi_{s}$ against position $x$ for a homogeneous
sample of target atoms. From left to right, the curves in cyan,
magenta, red, and blue refer to $\delta_{eff}/2\pi=-5.0$ MHz, $-1.2$
MHz, $-0.4$ MHz, and $-0.2$ MHz in order. Other parameters are
$N_{0}=2.0\times 10^{13}$ cm$^{-3}$, $L=30$ $\mu$m, and $x_{0}=-10$
$\mu$m except those specified at the beginning of sect.~III.}
\label{fig:Fig2}
\end{figure}

\vspace{2mm}

For a homogeneous sample of target atoms, we plot in
Fig.~\ref{fig:Fig2} the dispersion ($\chi_{s}^{\prime}$) and
absorption ($\chi_{s}^{\prime\prime}$) responses against position
$x$ by taking a few specific values of effective detuning
$\delta_{eff}$. It is clear that $\chi_{s}^{\prime}$ and
$\chi_{s}^{\prime\prime}$ exhibit quite narrow spatial profiles and
more importantly are out of phase (manifesting as an odd and an even
profile, respectively) to a good approximation as $\delta_{eff}$ is
decreased to be less than $-5$ MHz. It is also clear that the
absorption and dispersion profiles tend to be wider in space and
become more deviated from their counterparts in the frequency domain
as $\delta_{eff}$ is increased to be larger than $-0.2$ MHz.
Moreover, we note that the dispersion and absorption profiles may
move outside of the atomic sample in the case of
$\delta_{eff}\lesssim -16$ MHz or $\delta_{eff}\gtrsim 0$ MHz. These
findings can be well understood by looking back at Eq.~(\ref{Eqn5}),
with which we can determine a common center
$x_{c}=x_{0}+(-C_{6}/\delta_{eff})^{1/6}$ by setting
$\delta_{eff}+\mathcal{V}_{6}\rho_{33}=0$ while two half-widths
$\delta
x_{\pm}=x_{0}-x_{c}+[-C_{6}/(\delta_{eff}\mp\gamma_{rg})]^{1/6}$ by
setting $\delta_{eff}+\mathcal{V}_{6}\rho_{33}=\pm\gamma_{rg}$ with
respect to $\chi_{s}^{\prime}$ and $\chi_{s}^{\prime\prime}$ in the
limit of $\rho_{33}\to 1$. The nonlinear dependences of $x_{c}$ and
$\delta x_{\pm}$ on $\delta_{eff}$ answer for why the dispersion and
absorption profiles move toward the left side ($x=0$), become much
narrower, and look more symmetric as $\delta_{eff}$ is decreased,
\textit{e.g.}, from $-0.2$ MHz to $-5$ MHz.

\begin{figure}[tb]
\centering \includegraphics[width=0.48\textwidth]{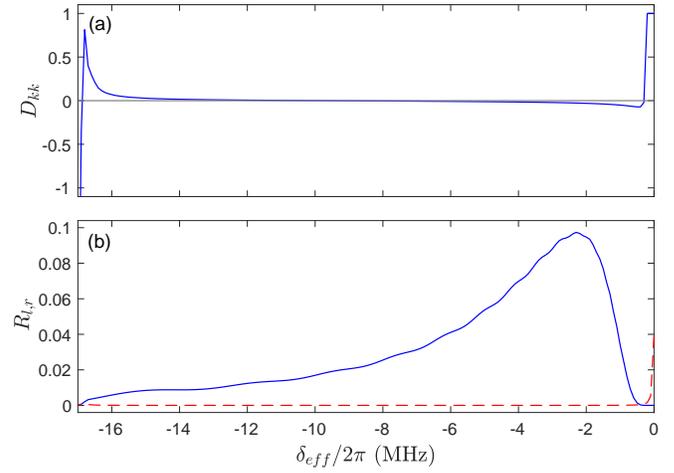}
\caption{(Color online) (a) Figure of merit $D_{kk}$ and (b)
reflectivities $R_{l,r}$ against effective detuning $\delta_{eff}$
for a homogeneous sample of target atoms. Relevant parameters are
the same as in Fig.~\ref{fig:Fig2} except $\lambda_{s}=780$ nm. The
blue-solid and red-dashed curves in (b) refer to $R_{l}$ and
$R_{r}$, respectively.} \label{fig:Fig3}
\end{figure}

Above results show that $\chi_{s}^{\prime}$ and
$\chi_{s}^{\prime\prime}$ generally don't vary in phase with the
increase or decrease of position $x$, hence are expected to satisfy
the spatial KK relation if both well contained in the finite atomic
sample. The fact is however that an essential part of the dispersion
and absorption profiles may extend outside of the finite atomic
sample when $\delta_{eff}$ is either two large or too small, leading
to a more or less violation of the spatial KK relation. This has
been evaluated by plotting figure of merit $D_{kk}$ in
Fig.~\ref{fig:Fig3}(a), from which we can see that the spatial KK
relation is roughly satisfied with $|D_{kk}|\leq 0.1$ in a wider
range between $\delta_{eff}/2\pi\lesssim -0.5$ MHz and
$\delta_{eff}/2\pi\gtrsim -16$ MHz, albeit well satisfied with
$|D_{kk}|\to 0$ in a narrower range centered at
$\delta_{eff}/2\pi\simeq -9$ MHz. Considering that spatial KK
relation is inseparable with unidirectional reflection, we further
plot in Fig.~\ref{fig:Fig3}(b) reflectivities $R_{l}$ and $R_{r}$
for a weak signal field incident from the left ($x=0$) and right
($x=L$) sides, respectively. It shows that unidirectional reflection
with $R_{l}\ne 0$ and $R_{r}\to 0$ occurs in the range between
$\delta_{eff}/2\pi\lesssim -0.5$ MHz and $\delta_{eff}/2\pi\gtrsim
-16.5$ MHz even if $|D_{kk}|$ has increased to be larger than $0.1$,
indicating that the spatial KK relation is not strictly required. It
is also worth noting that $R_{r}$ varies with $\delta_{eff}$ and
becomes maximal at $\delta_{eff}/2\pi\simeq-2.5$ MHz, in virtue of a
trade-off between the degree of spatial KK relation and the width of
real (dispersion) potential $\chi_{s}^{\prime}$.

\begin{figure}[tb]
\centering \includegraphics[width=0.48\textwidth]{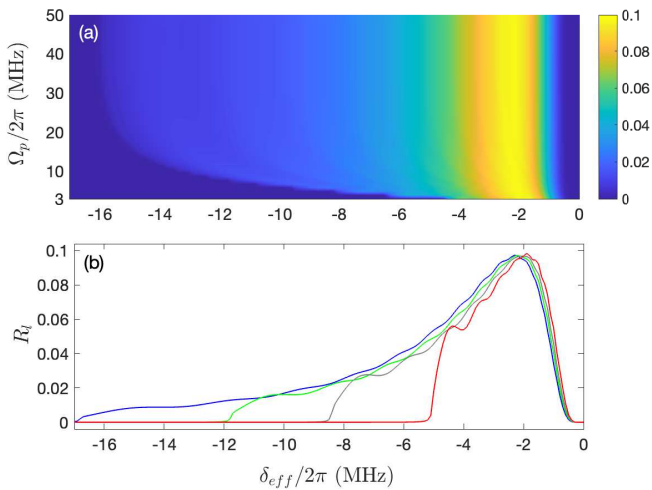}
\caption{(Color online) (a) Reflectivity $R_{l}$ against effective
detuning $\delta_{eff}$ and Rabi frequency $\Omega_{p}$ for a
homogeneous sample of target atoms with the same parameters as in
Fig.~\ref{fig:Fig3}. (b) 1D cuts of 2D plots in (a) with
$\Omega_{p}/2\pi=50$ MHz, $8.0$ MHz, $5.0$ MHz, and $3.0$ MHz from
left to right in order.} \label{fig:Fig4}
\end{figure}

\begin{figure}[tb]
\centering \includegraphics[width=0.48\textwidth]{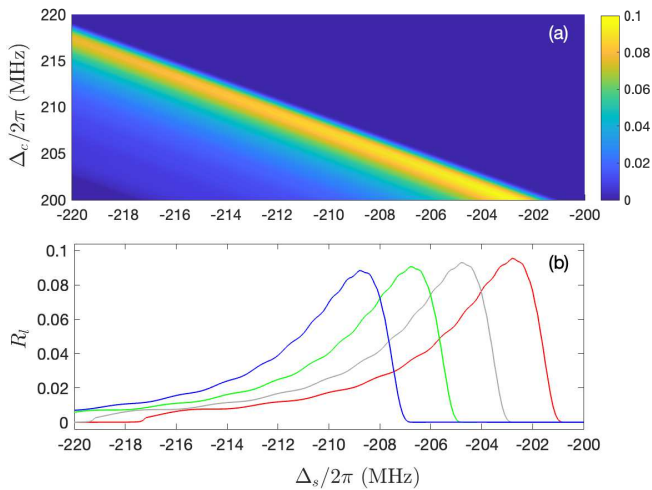}
\caption{(Color online) (a) Reflectivity $R_{l}$ against signal
detuning $\Delta_{s}$ and coupling detuning $\Delta_{c}$ for a
homogeneous sample of target atoms with the same parameters as in
Fig.~\ref{fig:Fig3}. (b) 1D cuts of 2D plots in (a) with
$\Delta_{c}/2\pi=206$ MHz, $204$ MHz, $202$ MHz, and $200$ MHz from
left to right in order.} \label{fig:Fig5}
\end{figure}

Then we examine two possibilities of dynamically modulating
unidirectional reflection behaviors based on nonlocal $vdW$
interactions between the control and target Rydberg atoms. One
possibility is shown in Fig.~\ref{fig:Fig4} where the pumping field
$\Omega_{p}$ is used as a remote `knob' to control the range of
$\delta_{eff}$ for observing unidirectional reflection. It is clear
that this range tends to be saturated in the case of
$\Omega_{p}\gtrsim 20$ MHz, but shrinks evidently from the side of
larger $|\delta_{eff}|$ as $\Omega_{p}$ gradually deviates from the
saturation regime. This can be attributed to the fact that a
decrease of $\Omega_{p}$ will result in a decrease of $\rho_{33}$
and thus a decrease of $x_{c}$ for a given $\delta_{eff}$,
equivalent to a decrease of the maximal $|\delta_{eff}|$ referring
to $x_{c}=0$ and denoting a boundary of the well satisfied spatial
KK relation. The other possibility is shown in Fig.~\ref{fig:Fig5}
where reflectivity $R_{l}$ is plotted against $\Delta_{s}$ instead
of $\delta_{eff}$, being $\Delta_{c}$ an alternative control `knob'.
It is easy to see that we can move the range of $\Delta_{s}$ for
observing unidirectional reflection as a whole, without shrinking or
expanding in terms of both $\Delta_{s}$ and $R_{r}$, by modulating
$\Delta_{c}$ in the limit of
$\Delta_{s}\simeq-\Delta_{c}\gg\Omega_{c}$. This fine tunability
relies on the fact that susceptibility $\chi_{s}$ in
Eq.~(\ref{Eqn5}) refers to an reduced two-level system where
effective detuning $\delta_{eff}$ is mainly contributed by the sum
of signal ($\Delta_{s}$) and coupling ($\Delta_{c}$) detunigns. A
reversed unidirectional reflection with $R_{l}=0$ and $R_{r}\ne 0$
can be attained by driving a second control atom at $x=L-x_{0}$ into
its Rydberg dark state while leaving the first control atom at
$x=x_{0}$ free of excitation (not shown).

\vspace{2mm}

So far we have shown that unidirectional reflection can be realized
and modulated for appropriate effective ($\delta_{eff}$) or signal
($\Delta_{s}$) detunings. However, the nonzero reflectivity
$R_{l}<0.1$ is obviously small because both real
($\chi_{s}^{\prime}$) and imaginary ($\chi_{s}^{\prime\prime}$)
potentials are rather weak (\textit{i.e.}, less than unit in
magnitudes). In order to enhance $\chi_{s}$ and thus increase
$R_{l}$, we can choose larger atomic density $N_{0}$ and/or smaller
dephasing rate $\gamma_{rg}$ as can be seen from Eq.~(\ref{Eqn5}).
Unfortunately, the former choice goes beyond the current
experimental technologies of cold atoms, while the latter choice is
restricted by the residual Doppler broadening of cold atoms
(\textit{e.g.}, $\sim20$ kHz at the temperature of $T=1$ $\mu$K).
This motivates us to consider another scenario where the homogeneous
atomic sample is replaced by a periodic atomic lattice described by
Eq.~(\ref{Eqn10}) so as to enhance the nonzero reflectivity by
incorporating multiple Bragg scattering into spatial KK relation.

\begin{figure}[tb]
\centering \includegraphics[width=0.48\textwidth]{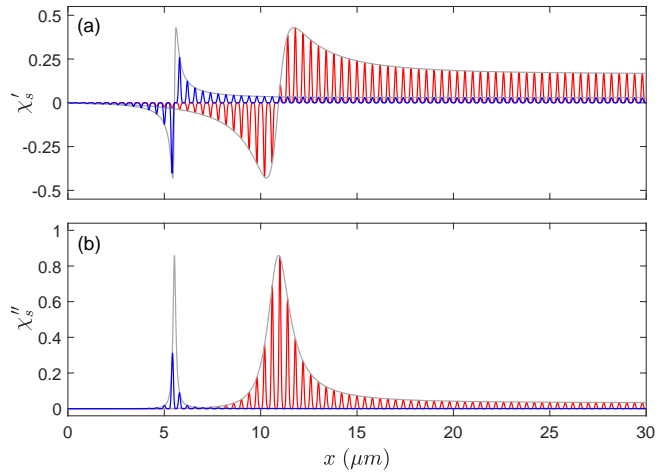}
\caption{(Color online) (a) Real and (b) imaginary parts of signal
susceptibility $\chi_{s}$ against position $x$ for a periodic
lattice of target atoms with $\delta_{eff}/2\pi=-1.2$ MHz (left) or
$-0.2$ MHz (right). Relevant parameters are the same as in
Fig.~\ref{fig:Fig2} except $\lambda_{s}=780$ nm, $\Lambda=400$ nm,
and $\delta x=\Lambda/6$.} \label{fig:Fig6}
\end{figure}

\begin{figure}[tb]
\centering \includegraphics[width=0.48\textwidth]{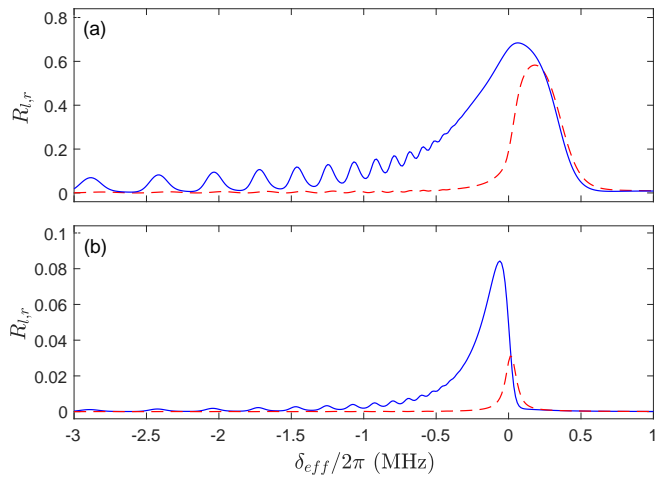}
\caption{(Color online) Reflectivities $R_{l}$ (blue-solid) and
$R_{r}$ (red-dashed) against effective detuning $\delta_{eff}$ for a
periodic lattice of target atoms with $N_{0}=2.0\times 10^{13}$
cm$^{-3}$ (a) or $2.0\times 10^{12}$ cm$^{-3}$ (b). Relevant
parameters are the same as in Fig.~\ref{fig:Fig2} except
$\lambda_{s}=780$ nm, $\Lambda=400$ nm, and $\delta x=\Lambda/6$.}
\label{fig:Fig7}
\end{figure}

Two typical examples on periodically modulated dispersion and
absorption responses are shown in Fig.~\ref{fig:Fig6} with
$\delta_{eff}/2\pi=-0.2$ MHz and $\delta_{eff}/2\pi=-1.2$ MHz,
respectively. It is easy to see that $\chi_{s}^{\prime}$ and
$\chi_{s}^{\prime\prime}$ are out of phase, to different extents
depending on $\delta_{eff}$, in their overall profiles similar to
their counterparts in Fig.~\ref{fig:Fig2}. But it is also obvious
that they exhibit comb-like fine structures under the not-in-phase
overall profiles as a result of the periodic Gaussian density $N(x)$
in Eq.~(\ref{Eqn10}). A signal field incident upon the finite atomic
lattice are expected to experience an enhanced reflection in the
presence of both spatial KK relation contributed by the overall
profiles of $\chi_{s}^{\prime}$ and $\chi_{s}^{\prime\prime}$ and
multiple Bragg scattering contributed by the fine structures of
$\chi_{s}^{\prime}$ and $\chi_{s}^{\prime\prime}$. This is exactly
what we observe in Fig.~\ref{fig:Fig7} where unbalanced
reflectivities $R_{l}$ and $R_{r}$ are plotted against effective
detuning $\delta_{eff}$.

We can see from Fig.~\ref{fig:Fig7}(a) that one reflectivity is
largely enhanced albeit in an asymmetric manner and exhibits a
maximum $R_{l}\to 0.7$ at $\delta_{eff}\simeq0$, while the other
reflectivity remains to be $R_{r}\to 0$ for $\delta_{eff}\lesssim
-0.5$ MHz. Moreover, it is worth noting that $R_{r}$ is also largely
enhanced for $\delta_{eff}\gtrsim 0$ and may even be equivalent to
$R_{l}$, indicating a fully destroyed spatial KK relation therein.
The underlying physics should be that strongest Bragg scattering
occurs around $\delta_{eff}\simeq 0$ where $\chi_{s}^{\prime}$ and
$\chi_{s}^{\prime\prime}$ exhibit very wide but not too low spatial
profiles on one hand and tend to vary in phase on the other hand.
Fig.~\ref{fig:Fig7}(b) further shows that the asymmetric enhancement
of $R_{l,r}$ holds for a smaller atomic density and more importantly
the maximal value $R_{l}\simeq 0.085$ in a periodic atomic lattice
could be equivalent to that in a homogeneous atomic sample with a
ten-times larger density $N_{0}$, see Fig.~\ref{fig:Fig3}(b). These
results confirm that multiple Bragg scattering is a valid tool for
improving the asymmetric or unidirectional reflection behaviors
arising from spatial KK relation, which is unattainable yet by
inserting a homogeneous atomic sample into a Fabry-Perot cavity (not
shown).

\bigskip

\section{Conclusions}
In summary, we have proposed an efficient scheme for realizing the
dynamically tunable spatial KK relation by exploiting nonlocal $vdW$
interactions of Rydberg atoms. One control atom in a Rydberg dark
state is used to map the dispersion and absorption responses of a
homogenous sample or a periodic lattice of target atoms from the
frequency domain to the space domain. This is attained as all target
atoms are driven in the EIT regime to an effective two-level
configuration by a signal and a coupling field kept near resonance
on one two-photon transition but far-detuned from two single-photon
transitions. Our numerical results show that the spatial dispersion
and absorption responses generally don't vary in phase and more
importantly could well satisfy the spatial KK relation, hence
supporting unidirectional ($R_{l}\ne 0$ and $R_{r}=0$) reflection
behaviors. Note also that periodic atomic lattices seem more
appealing than homogenous atomic samples in that they promise an
obvious enhancement of the nonzero reflection due to a positive
interplay of multiple Bragg scattering and spatial KK relation. Our
findings should be instructive on combining non-Hermitian quantum
optics and coherent manipulation of Rydberg atoms, \textit{e.g.}, to
develop one-way optical devices and explore new applications with
long-range $vdW$ interactions.

\section*{Acknowledgement}
Supported by National Natural Science Foundation of China
(No.~12074061), Funding from Ministry of Science and Technology of
China (No.~2021YFE0193500), and Science Foundation of Education
Department of Jilin Province (No.~JJKH20211279KJ).

\end{document}